\documentclass[pra,twocolumn]{revtex4-2}
\usepackage{url}
\usepackage{amsfonts}
\usepackage[T1]{fontenc}
\usepackage{float}
\usepackage[utf8]{inputenc}
\usepackage{epsfig,epsf}
\usepackage{subcaption}
\usepackage{amsbsy}
\usepackage{amsfonts}
\usepackage{textcomp}
\usepackage{amssymb}
\usepackage{mathrsfs}
\usepackage{graphicx}
\usepackage{amsmath}
\usepackage{latexsym}
\usepackage{xcolor}
\usepackage{color}

\begin{document}

\title{Generalized Schwarz and Jensen  inequalities and the resulting uncertainty relations\\ and correlations}
\author{Krzysztof Apolinary  Urbanowski}
\affiliation{Institute of Physics,
University of Zielona G\'{o}ra,  \\
ul. Prof. Z. Szafrana 4a,
65-516 Zielona G\'{o}ra,
Poland}
\email[]{K.Urbanowski@if.uz.zgora.pl, $\;$ k.a.urbanowski@gmail.com}

\date{\today}
\newtheorem{theorem}{Theorem}

\begin{abstract}
We analyze the Schwarz inequality and its generalizations, as well as inequalities resulting from the Jensen inequality. They  are used in quantum theory to derive the Heisenberg--Robertson (HR) and Schr\"{o}\-dinger--Robertson (SR) uncertainty relation for two non--commuting observables and their generalizations to three or more non--commuting observables. Jensen's inequality, in turn, is helpful in deriving various the  "sum uncertainty relations" for two or more observables.
Using these inequalities, we derive various types of generalized uncertainty relations for more than two non--commuting observables and analyze their properties and critical points.
We also study the connections between the generalizations of the HR and SR uncertainty relations for two and more observables and the correlations of these observables in the state of the quantum system under study.
In this analysis, we pay special attention to the consequences of the generalized SR uncertainty relation for three non--commuting observables on their correlations in a given state of a quantum system and to the connections of this relation with the appropriate correlation matrix, whose matrix elements are the quantum versions of the Pearson coefficient.
We show also that the SR uncertainty relation  (including the generalized ones) can be written in an equivalent way using  these Pearson coefficients.

\end{abstract}

\maketitle

\section{Introduction}

Quantum theory describing phenomena occurring in the quantum world results from several basic assumptions (see e.g. \cite{Gal}).
One of them states that each state of a quantum system is uniquely assigned a normalized vector in a vector  space with an inner product, i.e., a Hilbert space ${\cal H}$, which is called the state space.
Another assumption says that there is a unique correspondence between measurable physical quantities, say $F$,  and Hermitian operators acting in the space ${\cal H}$.
These quantities are called observables and operators represented them mathematically act on the quantum state, say $|\psi\rangle \in {\cal H}$, of the system to produce real--valued eigenvalues, which are possible outcomes of a measurement.
If now the system is in state $|\psi_{f}^{F}\rangle$ which is an eigenstate of operator $F$ then the result of measuring the observable $F$ in this state will be the eigenvalue $f$ of the operator $F$: $F|\psi_{f}^{F}\rangle = f |\psi_{f}^{F}\rangle$. Then simply the expectation value of observable $F = \langle \psi_{f}^{F}|F|\psi_{f}^{F}\rangle$ will be equal $f$. (Here, for simplicity, it was assumed that $f$ belongs to the set of discrete eigenvalues of $F$).
One of the important problems in quantum theory is the question what will be the result of measuring the mentioned observable $F$ when the quantum system is in state $|\phi\rangle \in {\cal H}$ which is not an eigenvector of operator $F$.
In such a case, quantum mechanics cannot predict the exact result of a future measurement. Simply quantum mechanics only says  that the result of such a measurement should be one of the eigenvalues of the the operator representing $F$, but it does not say which one ---  it only determines the possible probability of such an outcome.
In general,
outcome are random and are distributed with the Born rule \cite{Bes,BM1,BM2}.
So, if we plan to perform many such measurements, we can only predict the probable average (or expected) value of the measured quantity $F$ in the state $|\phi\rangle$: $\langle F\rangle_{\phi} = \langle \phi |F |\phi \rangle$
provided that $|\langle\phi|F|\phi \rangle |< \infty$.
We can now ask what the average discrepancy between the result of a single measurement of a given quantity in state $|\phi\rangle$ and the expected $\langle F\rangle_{\phi} $ value might be.
Mathematically, this discrepancy, $\Delta_{\phi}F$,  called the "standard deviation," is described by the following formula:
\begin{equation}
\Delta_{\phi} F = \| \delta_{\phi} F|\phi\rangle\| \geq 0, \label{DF}
\end{equation}
where
\begin{equation}
\delta_{\phi} F |\phi\rangle = (F - \langle F\rangle_{\phi}\,\mathbb{I})|\phi\rangle. \label{dF}
 \end{equation}
 Equivalently:  $\Delta_{\phi} F \equiv \sqrt{\langle F^{2}\rangle_{\phi} - \langle F\rangle_{\phi}^{2}}$.

The problem becomes more complicated when we plan to measure simultaneously two observables represented by non--commuting operators $A$ and $B$ in the same state $|\phi\rangle$ of the system.
In such a situation, observable $A$ affects the possible results of the measurement of observable $B$ disturbing them, and vice versa.
The question arises whether there is any relationship between the standard deviations, $\Delta_{\phi}A$ and $\Delta_{\phi}B$,  of observables $A$ and $B$ in this situation, and if so, how to describe it.
This problem explains
the famous uncertainty principle discovered by Heisenberg \cite{H,H2,Robertson}, (see also \cite{Dod,Con}). This principle belongs to the one of characteristic and the most important consequences of the quantum mechanics and indicates the fundamental differences between classical and quantum world.
 Probably the most common form of this uncertainty principle is
 \begin{equation}
\Delta_{\phi} A \cdot \Delta_{\phi} B\;\geq\;\frac{1}{2} \left|\langle [A,B] \rangle_{\phi} \right|,\label{R1}
\end{equation}
which holds for any two observables, $A$ and $B$, represented by non--commuting hermitian operators $A$ and $B$ acting in ${\cal H}$ (see \cite{Robertson} and also \cite{M,Teschl}), such that $[A,B]$ exists and $|\phi\rangle \in {\cal D}(AB) \bigcap {\cal D}(BA)$, (${\cal D}({\cal O})$ denotes the domain of an operator $\cal O$ or of a product of operators).
The derivation of  inequality (\ref{R1}) is the rigorous one: It is a simple consequence of  the Schwarz inequality.

So,
we can say that the relation (\ref{R1}) results from basic assumptions of the quantum theory and from the geometry of Hilbert space \cite{Gal,M,Teschl}.
The uncertainty principle of the form (\ref{R1}) was generalized to a more general form by Schr\"{o}dinger (see \cite{Schrod-1930}).

The geometry of vectors in the Hilbert space of states of a quantum system was used as a tool to find the so--called "sum uncertainty relation" \cite{Pat} and "stronger sum uncertainty relation" \cite{Mac} and  related  relations. The sum uncertainty relation results from the triangle inequality and has the following form:
\begin{equation}
\Delta_{\phi}A + \Delta_{\phi}B \geq \Delta_{\phi}(A + B). \label{Pat1}
\end{equation}
The stronger uncertainty relation results from the triangle inequality of the second kind \cite{Dad,Hsu} (i.e. the Jensen inequality for squares of norms of two vectors \cite{cv,mk}) and it looks as follows,
\begin{equation}
(\Delta_{\phi}A)^{2} + (\Delta_{\phi}B)^{2} \geq	 \frac{1}{2} \left(\Delta_{\phi}(A+B)\right)^{2}, \label{M12}
\end{equation}
In all the above inequalities, the standard deviations are defined by the relation (\ref{DF}).

The problem becomes more complicated when one wants to predict and estimate the possible outcome of simultaneous measurement of several, more than two, observables described by non--commuting operators.
This can be done using, for example, the generalization of the triangle inequality to $N$ vectors  or other inequalities relating the norms of $N$ vectors, where $N \geq 3$.
Some attempts to solve this problem can be found, for example, in \cite{Pat,Bin}, where the generalized triangle inequality was used, or in \cite{Bin,Xiao,Son}, where the generalization of relation (\ref{M12}) to $N$ observables was discussed.
The problem with various generalizations of this type is that they often lead to uncertainty principles that have a rather complicated form and are therefore difficult to apply to specific problems.
This inspired some of the reflections contained in this paper.
The goal is to exploit certain inequalities that hold in normed spaces and inner product spaces, and using them, find possibly simple generalizations of the aforementioned uncertainty principles, and their analysis.

The paper is organized as follows:
Section 2 introduces readers to some inequalities in an inner product space and their generalizations, which may be useful in deriving the uncertainty relations and their generalizations.
In Section 3, the inequalities discussed in Sec. 2 are used to derive the  HR  uncertainty relation (\ref{R1})  and the RS uncertainty relation for two non--commuting observables and their generalizations to the case of several non--commuting observables.
Section 4 contains a derivation of the sum uncertainty relations (\ref{Pat1}), (\ref{M12}) and their  generalizations to the case of more than two nono--commuting obserwables. In Section 5 critical points of the sum uncertainty relations,  HR and SR uncertainty relations are analyzed. In Section 6 we analyze connections of SR and generalized SR uncertainty relations with correlation function and Pearson coefficient.
Particular attention is paid here to the case of the uncertainty relation for three non--commuting observables and the connection of this relation with the correlation function for a quantum system in which such observables act.
Concluding remarks can be found in Sec. 7

\section{Mathematical tools: useful inequalities}

As mentioned in the previous Section, a rigorous derivation of lower bounds on the product of standard deviations or variances, as well as on their sums, requires the use of certain inequalities for the norms of vectors in the Hilbert state space ${\cal H}$. Below we discuss the inequalities that will be needed later.
The starting point for deriving the Heisenberg--Robertson (HR) and Robertson--Schrodinger (SR) uncertainty relations is usually the Cauchy--Schwarz (CS) inequality \cite{Teschl,Schrod-1930},
\begin{equation}
\left\|\;|\psi_{1}\rangle \right\| \cdot\left\|\;|\psi_{2}\rangle\right\| \geq \left| \langle \psi_{1}|\psi_{2} \rangle \right|, \label{S}
\end{equation}
where $|\psi_{1}\rangle, |\psi_{2}\rangle \in {\cal H}$. If one now wants to extend the HR uncertainty relation to three or more non-commuting observables, one must find a generalization of the CS inequality to more than two vectors. Let us consider the case of three vectors $|\psi_{1}\rangle, |\psi_{2}\rangle, |\psi_{3}\rangle \in {\cal H}$.
For these vectors, three CS inequalities  (\ref{S})  can be written,
\begin{eqnarray}
\left\|\;|\psi_{1}\rangle \right\| \cdot \left\|\;|\psi_{2}\rangle\right\| &\geq & \left| \langle \psi_{1}|\psi_{2} \rangle \right|, \nonumber \\
\left\|\;|\psi_{2}\rangle \right\| \cdot \left\|\;|\psi_{3}\rangle\right\| &\geq &\left| \langle \psi_{2}|\psi_{3} \rangle \right|, \nonumber \\
\left\|\;|\psi_{1}\rangle \right\| \cdot \left\|\;|\psi_{3}\rangle\right\| &\geq &\left| \langle \psi_{1}|\psi_{3} \rangle \right|. \label{3S}
\end{eqnarray}
Then, multiplying all these inequalities side by side, we get
\begin{equation}
\left\|\;|\psi_{1}\rangle \right\|^{2} \cdot \left\|\;|\psi_{2}\rangle\right\|^{2} \cdot \left\|\;|\psi_{3}\rangle\right\|^{2}
\geq  \left| \langle \psi_{1}|\psi_{2} \rangle \right| \, \left| \langle \psi_{2}|\psi_{3} \rangle \right| \,\left| \langle \psi_{1}|\psi_{3} \rangle \right|. \label{S3}
\end{equation}
Similarly, we can generalize the inequality (\ref{S}) to four vectors $|\psi_{1}\rangle, |\psi_{2}\rangle,  |\psi_{3}\rangle,  |\psi_{4}\rangle  \in {\cal H}$.
We simply need to write Cauchy--Schwarz's inequalities (\ref{S}) for all possible pairs of these vectors and then multiply all these inequalities side by side respectively. The result is
\begin{align}
\left\|\;|\psi_{1}\rangle \right\|^{3} \cdot \left\|\;|\psi_{2}\rangle\right\|^{3}  \cdot & \left\|\;|\psi_{3}\rangle\right\|^{3} \cdot \left\|\;|\psi_{4}\rangle\right\|^{3} \nonumber \\
& \geq  \left| \langle \psi_{1}|\psi_{2} \rangle \right| \, \left| \langle \psi_{1}|\psi_{3} \rangle \right| \,\left| \langle \psi_{1}|\psi_{4} \rangle \right| 
\nonumber \\
& \times  
 \left| \langle \psi_{2}|\psi_{3} \rangle \right| \, \left| \langle \psi_{2}|\psi_{4} \rangle \right| \,\left| \langle \psi_{3}|\psi_{4} \rangle \right|. \label{S4}
\end{align}
In a similar way one can generalize the inequality (\ref{S}) to  the case of five and more vectors (see, e. g.  \cite{Had}).

In the literature, one can find other generalizations and extensions of the CS inequality to three vectors, which can be used to generalize the HR and RS uncertainty relations to the case of three non--commuting observables.
 The Buzano inequality is such a generalization \cite{Buz,ar,Lu,Min}.  It has the following form
 \begin{equation}
 \frac{1}{2}\left(\left\| \,\psi_{1}\rangle \right\| \,\left\| \,|\psi_{2}\rangle \right\| +
 \left|\langle\psi_{1}|\psi_{2}\rangle \right|\right)\, \left\|\,|\psi_{3}\rangle\right\|^{2} \geq  \left| \langle \psi_{1}|\psi_{3}\rangle \langle \psi_{3}|\psi_{2}\rangle \right|.  \label{B1}
 \end{equation}

 The related and useful generalization of this inequality is the following one:
 \begin{multline}
 \left\| \,|\psi_{1}\rangle \right\|^{2}\, \left\| \,|\psi_{2}\rangle \right\|^{2} \, \left\|\,|\psi_{3}\rangle\right\|^{2} + \\ 2
  \left| \langle\psi_{1}|\psi_{2}\rangle  \langle \psi_{2}|\psi_{3}\rangle  \langle \psi_{3}|\psi_{1}\rangle \right| \geq \\
 \geq  \left\|\;|\psi_{1} \rangle \right\|^{2} \, \left|  \langle \psi_{2}|\psi_{3} \rangle \right|^{2}
 +  \left\|\;|\psi_{2} \rangle \right\|^{2} \, \left|  \langle \psi_{3}|\psi_{1} \rangle \right|^{2}\\
 +   \left\|\;|\psi_{3} \rangle \right\|^{2} \, \left| \langle \psi_{1}|\psi_{2} \rangle \right|^{2}, \label{Lu1}
 \end{multline}
and it is a generalized version of the inequality considered in  \cite{sh1} and latter in  \cite{ar}. This generalization has been refreshed and discussed by Cezar Lupu and Dan Schwarz \cite{Lu} and then also considered, e.g. in \cite{Min,Min2}.

Note that if we use the property that the modulus of the inner product  $ \left| \langle\psi_{1}|\psi_{2}\rangle \right|$ on the left side of inequality (\ref{B1}) is smaller than the product of the norms of these vectors (see Eq. (\ref{S})), we obtain the less restrictive version of Buzano's inequality (\ref{B1}),
\begin{equation}
\left\| \,|\psi_{1}\rangle \right\| \,\left\| \,|\psi_{2}\rangle \right\| \, \left\|\,|\psi_{3}\rangle\right\|^{2} \geq  \left| \langle \psi_{1}|\psi_{3}\rangle \langle \psi_{3}|\psi_{2}\rangle \right|. \label{B2}
\end{equation}
On the other hand, the same inequality is obtained if we multiply the second and third inequalities in (\ref{3S}) by each other, respectively. Thus, the weaker version of Buzano's inequality is equivalent to the product of two appropriately chosen CS inequalities.

Similarly,  using the CS inequality (\ref{S}) we can obtain the weaker version the
inequality (\ref{Lu1}),
\begin{multline}
3 \,\left\| \,|\psi_{1}\rangle \right\|^{2}\, \left\| \,|\psi_{2}\rangle \right\|^{2} \, \left\|\,|\psi_{3}\rangle\right\|^{2} \geq
 \left\|\;|\psi_{1} \rangle \right\|^{2} \,  \left|  \langle \psi_{2}|\psi_{3} \rangle \right|^{2} \\
 +   \left\|\;|\psi_{2} \rangle \right\|^{2} \, \left| \langle \psi_{3}|\psi_{1} \rangle \right|^{2}
 +  \left\|\;|\psi_{3} \rangle \right\|^{2} \, \left|  \langle \psi_{1}|\psi_{2} \rangle \right|^{2}. \label{Lu2}
\end{multline}


The uncertainty relation (\ref{M12}) for the sum of variances can be easily generalized to the case of N non--commuting observables if one uses the observation that the norm of vectors is a convex function.
Recall that for the function $f:X \longrightarrow \mathbb{R}$ we will say that it is convex on the set $X$ if for any $x,y \in X$ and any $\alpha \in (0,1)$ we have (see, e.g \cite{cv,cp}),
\begin{equation}
f(\alpha x + (1 - \alpha y)) \leq
\alpha f(x) + (1 - \alpha) f (y). \label{conv}
\end{equation}
If $f:X \longrightarrow \mathbb{R}$ is a continuous function  then it is convex if and only if it is midpoint convex, that is, if
\begin{equation}
f\left(\frac{x +  y}{2}\right) \leq
\frac{f(x) +  f (y)}{2}, \label{mid-con}
\end{equation}
for all $x,y \in X$ (see, eg. \cite{cp}).
These properties are important for our considerations, because,  as is easy to prove, the functions  $f: \phi \in {\cal X} \longrightarrow \|\,\phi\| \in \mathbb{R} $ and $f: \phi \in {\cal X} \longrightarrow \|\,\phi\|^{2} \in \mathbb{R}$ are convex  for any norm in a normed space  $\cal X$. For convex functions, Jensen's inequality holds,
\begin{equation}
f\left(\sum_{i=1}^{n} p_{i} \phi_{i}\right) \leq \sum_{i=1}^{n} p_{i} f(\phi_{i}), \label{J1}
\end{equation}
where $f$ is convex function on $\cal X$, $f: {\cal X} \longrightarrow  \in \mathbb{R} $,  $n \in \mathbb{N}$, and  $p_{1},p_{2}, \ldots, p_{n} \in (0,1)$ are such that  $\sum_{i=1}^{n} p_{i} = 1$. This inequality holds for any $\phi_{1}, \phi_{2}, \ldots, \phi_{n} \in{\cal X}$.
If we now apply Jensen's inequality to the function $ f(|\psi\rangle) = \| \,\psi\rangle \|$, we get that
\begin{equation}
\|\,\sum_{i=1}^{n} p_{i} |\psi_{i}\rangle \| \leq \sum_{i=1}^{n} p_{i} \|\,|\psi_{i}\rangle \|, \label{J2}
\end{equation}
where $|\psi_{i}\rangle \in {\cal H}$, $p_{1},p_{2}, \ldots, p_{n} \in (0,1)$ and $\sum_{i=1}^{n} p_{i} = 1$. In turn, choosing $ f(|\psi\rangle) = \| \,\psi\rangle \|^{2}$,  for the same $p_{i}$ and $|\psi_{i}\rangle \in {\cal H}$, we get
\begin{equation}
\|\,\sum_{i=1}^{n} p_{i} |\psi_{i}\rangle \|^{2} \leq \sum_{i=1}^{n} p_{i} \|\,|\psi_{i}\rangle \|^{2}. \label{J3}
\end{equation}
Further, putting in (\ref{J2}), e.g. $p_{i} = \frac{1}{n}$, $n \geq 2$, we get that
\begin{equation}
 \sum_{i=1}^{n} \|\,|\psi_{i}\rangle \| \geq \|\,\sum_{i=1}^{n} |\psi_{i}\rangle \|, \label{J2a}
\end{equation}
which is the generalized triangle inequality. In turn, substituting $p_{i} = \frac{1}{n}$  Eq. (\ref{J3}) gives
\begin{equation}
 \sum_{i=1}^{n} \|\,|\psi_{i}\rangle \|^{2} \geq  \frac{1}{n}\|\,\sum_{i=1}^{n} |\psi_{i}\rangle \|^{2}. \label{J3a}
\end{equation}

Let us note that from (\ref{J3a}) we obtain for $n=2$ the following inequality
\begin{equation}
\|\,|\psi_{1}\rangle \|^{2} + \|\,|\psi_{1}\rangle \|^{2} \geq \frac{1}{2} \|\,|\psi_{1}\rangle  + |\psi_{2}\rangle\|^{2}. \label{J3b}
\end{equation}
This is Jensen's inequality for two variables, also known as the triangle inequality of the second kind  \cite{Hsu}.

The properties of the inequalities described and derived above will be used in the subsequent parts of the paper to find generalizations of the uncertainty relations mentioned in Sec. 1.

\section{Generalizations of the \\Heisenberg--Robertson \\ and Robertson--Schr\"{o}dinger\\ uncertainty relations}

We will now use the inequalities discussed in the previous Section to find the standard and also some new uncertainty relations.
Let us consider the set of non-commuting observables $\{A_{i}\}_{i=1}^{N}$, $[A_{i},A_{j}] \neq 0$ for $i \neq j$ and define
\begin{equation}
|\psi_{i}\rangle = \delta_{\phi} A_{i}|\phi\rangle, \;\;(i=1,2,\;\dots,N),\label{dAi}
\end{equation}
where $\delta_{\phi}A_{i} |\phi\rangle$ is defined by (\ref{dF}).
Let us now assume that \linebreak $|\phi\rangle \in  \bigcap_{i}^{N}\left[\bigcap_{j=1,j\neq i}^{N} \left({\cal D}(A_{i}A_{j})\right)\right]$,  (where ${\cal D}(O)$ is a domain of the operator $O$, or a product of operators),
and  substitute (\ref{dAi})  into (\ref{S}). As a result, the CS inequality (\ref{S}) becomes the uncertainty relation SR after using equation (\ref{DF}):
\begin{equation}
 (\Delta_{\phi} A_{i} )\, \cdot\,( \Delta_{\phi} A_{j})\,\geq |{\cal C}_{\phi}(A_{i},A_{j})|, \;\; (i \neq j), \label{RS2}
\end{equation}
where,
\begin{multline}
{\cal C}_{\phi}(A_{i},A_{j}) = \langle \psi{_i}|\psi_{j}\rangle \equiv \langle \phi| \delta_{\phi} A_{i}\;\delta_{\phi} A_{j}|\phi\rangle\\
 \equiv \langle A_{i}A_{j} \rangle_{\phi} - \langle A_{i}\rangle_{\phi}\,\langle A_{j}\rangle_{\phi}, \label{C}
\end{multline}
 is a quantum version of the correlation function (or  covariance). Here defining the correlation function ${\cal C}_{\phi}(A_{i},A_{j})$ we follow, e. g.  \cite{Poz,Khr} and others.
Since $|{\cal C}_{\phi}(A_{i}A_{j})| \geq |\Im \,[{\cal C}_{\phi}(A_{i}A_{j})]| \equiv \frac{1}{2} \left| \langle [A_{i},A_{j}]\rangle_{\phi} \right|$, (where $\Im\, [z]$ denotes the imaginary part of $z$), the HR uncertainty relation immediately  follows from (\ref{RS2}):
\begin{equation}
(\Delta_{\phi} A_{i} )\, \cdot\,( \Delta_{\phi} A_{j})\,\geq \frac{1}{2} \left|\langle[A_{i},A_{j}]\rangle_{\phi}\right|, \;(i \neq j). \label{HR2}
\end{equation}
Relations (\ref{RS2}) and (\ref{HR2})  are sometimes called "standard uncertainty relations".

An analogous method applied to the inequality (\ref{S3})  gives the following results:
\begin{multline}
(\Delta_{\phi} A_{1} )^{2}\, \cdot\,( \Delta_{\phi} A_{2})^{2}\,\cdot\, ( \Delta_{\phi} A_{3})^{2}\geq \\
\geq |{\cal C}_{\phi}(A_{1},A_{2})|\,|{\cal C}_{\phi}(A_{2},A_{3})|\,|{\cal C}_{\phi}(A_{1},A_{3})|,\label{RS3}
\end{multline}
which is a generalization of the RS uncertainty relation  to the case of three non--commuting observables $A_{1}, A_{2}, A_{3}$.  The generalization of the HR relation to the case of three non--commuting observables can be obtained from the inequality (\ref{RS3}) similarly to (\ref{HR2}) from (\ref{RS2}):
\begin{multline}
(\Delta_{\phi} A_{1} )^{2}\, \cdot\,( \Delta_{\phi} A_{2})^{2}\,\cdot\, ( \Delta_{\phi} A_{3})^{2}\geq \frac{1}{8}  \left|\langle [A_{1},A_{2}]\rangle_{\phi}\right| \cdot \\ \cdot \left|\langle [A_{2},A_{3}]\rangle_{\phi}\right|  \cdot \left|\langle [A_{1},A_{3}]\rangle_{\phi}\right|.\label{HR3}
\end{multline}

In the case of four non--commuting observables $A_{1}, A_{2}, A_{3}, A_{4}$, using (\ref{dAi}) and (\ref{DF}) we obtain from (\ref{S4}) the following following generalization of the uncertainty relation RS,
\begin{align}
(\Delta_{\phi} A_{1} )^{3} \cdot ( \Delta_{\phi} A_{2})^{3} \cdot & ( \Delta_{\phi} A_{3})^{3} \cdot ( \Delta_{\phi} A_{4})^{3}  \geq  \nonumber \\
& \geq  |{\cal C}_{\phi}(A_{1},A_{2})|\cdot |{\cal C}_{\phi}(A_{1},A_{3})| \nonumber \\
& \cdot |{\cal C}_{\phi}(A_{1},A_{4})|\cdot|{\cal C}_{\phi}(A_{2},A_{3})| \nonumber \\
& \cdot  |{\cal C}_{\phi}(A_{2},A_{4})|\cdot|{\cal C}_{\phi}(A_{3},A_{4})|. \label{RS4}
\end{align}
The generalized HR uncertainty relation in this case looks as follows:
\begin{align}
(\Delta_{\phi} A_{1} )^{3} \cdot ( \Delta_{\phi} A_{2})^{3} \cdot &  ( \Delta_{\phi} A_{3})^{3} \cdot ( \Delta_{\phi} A_{4})^{3}  \geq \nonumber \\
& \geq \frac{1}{16}  \left|\langle [A_{1},A_{2}]\rangle_{\phi}\right| \cdot \left|\langle [A_{1},A_{3}]\rangle_{\phi}\right|  \nonumber \\
& \cdot \left|\langle [A_{1},A_{4}]\rangle_{\phi}\right| \cdot \left|\langle [A_{2},A_{3}]\rangle_{\phi}\right|  \nonumber \\
& \cdot  \left|\langle [A_{2},A_{4}]\rangle_{\phi}\right|  \cdot \left|\langle [A_{3},A_{4}]\rangle_{\phi}\right|.
\label{HR4}
\end{align}
In a similar way, one can generalize the uncertainty relations HR and RS to the case of $N \geq 5$  non--commuting observables (see also, \cite{Had,Xiao1}.

In turn, starting from the Buzano inequality (\ref{B1}), and similarly to the relations obtained above, using (\ref{dAi}) and (\ref{DF}) we obtain another version of the RS uncertainty principle for three non--commuting observables, $A_{1}, A_{2}, A_{3}$,
\begin{align}
\left( \Delta_{\phi}A_{1}\cdot \Delta_{\phi}A_{2} \right. & + \left.| {\cal C}_{\phi}(A_{1} A_{2}) | \right)  (\Delta_{\phi}A_{3})^{2} \geq \nonumber \\
& \geq 2 \left|{\cal C}_{\phi}(A_{1} A_{3})\,{\cal C}_{\phi}(A_{3} A_{2})\right|. \label{B1a}
\end{align}
If we use a less restrictive version of Buzano's inequality (\ref{B2}), we obtain the following generalization of the RS uncertainty principle instead of (\ref{B1a}),
\begin{equation}
\Delta_{\phi}A_{1}\cdot \Delta_{\phi}A_{2}\cdot (\Delta_{\phi}A_{3})^{2} \geq \left|{\cal C}_{\phi}(A_{1} A_{3})\,{\cal C}_{\phi}(A_{3} A_{2})\right|. \label{B2a}
\end{equation}
This gives us another generalization of the HR uncertainty relation for three non--commuting observables,
\begin{equation}
\Delta_{\phi}A_{1}\cdot \Delta_{\phi}A_{2}\cdot (\Delta_{\phi}A_{3})^{2} \geq \frac{1}{4} \left|\langle [A_{1},A_{3}]\rangle_{\phi}\right| \cdot \left|\langle [A_{3},A_{2}]\rangle_{\phi}\right|. \label{B2b}
\end{equation}
One can expect that generalizations of uncertainty relations  (\ref{B2a}) and (\ref{B2b}) will be much easier in practical applications than relation (\ref{B1a}).
However, some problems may arise when using these generalizations as well as the relation (\ref{B1a}).
This is due to the asymmetric contribution of the observables $A_{1}, A_{2}, A_{3}$ to these relations.
This drawback can be eliminated if, for example, together with relation (\ref{B2a}), to consider two new relations obtained by swapping the positions of observables $A_{2}$ and $A_{3}$ in (\ref{B2a}), (this will be the first new relation), and $A_{1}$ and $A_{3}$ in (\ref{B2a}), (which will give us the second new relation).
The next step is to multiply these two new inequalities and inequality (\ref{B2a}) by their respective sides. The new inequality thus obtained does not have the above--mentioned drawback, but it is similar to (\ref{HR3}).

Let us see now what are implications of using of the
inequality (\ref{Lu1}) to derive a generalization of the RS uncertainty relation for thee case of three non--commuting observables  $A_{1}, A_{2}, A_{3}$.
To do this we should use inequality (\ref{Lu1}) and repeat  all the steps leading to above derived generalizations of the RS and HR uncertainty relations.
Therefore, we should substitute the vectors $|\psi_{1}\rangle, |\psi_{2}\rangle, |\psi_{3}\rangle$  into (\ref{Lu1}) in the form (\ref{dAi}) and use definitions (\ref{DF}) and (\ref{C}). The result is
\begin{align}
(\Delta_{\phi}A_{1})^{2} & \cdot (\Delta_{\phi}A_{2})^{2} \cdot (\Delta_{\phi}A_{3})^{2}  \geq   (\Delta_{\phi}A_{1})^{2}\,|{\cal C}_{\phi}(A_{2} A_{3})|^{2} \nonumber \\
& + (\Delta_{\phi}A_{2})^{2}\,|{\cal C}_{\phi}(A_{3} A_{1})|^{2} \nonumber  \\
& + (\Delta_{\phi}A_{3})^{2}\,|{\cal C}_{\phi}(A_{1} A_{2})|^{2} \nonumber  \\
& - \,2\,\left|{\cal C}_{\phi}(A_{1} A_{2})\,{\cal C}_{\phi}(A_{2} A_{3})\,{\cal C}_{\phi}(A_{1} A_{3})\right|. \;\;\;\;\; \label{Lu1a}
\end{align}
In turn, repeating this procedure in the case of a less restrictive version of the
inequality (\ref{Lu2}) leads to the following result
\begin{align}
(\Delta_{\phi}A_{1})^{2} \cdot (\Delta_{\phi}A_{2})^{2} &\cdot (\Delta_{\phi}A_{3})^{2} \geq \frac{1}{3} \left[\, (\Delta_{\phi}A_{1} )^{2}\,| {\cal C}_{\phi} (A_{2} A_{3}) |^{2} \right. \nonumber \\
&+ \;\;(\Delta_{\phi}A_{2})^{2}\,|{\cal C }_{\phi}(A_{3} A_{1})|^{2}   \nonumber \\
&  + \;\left.  (\Delta_{\phi}A_{3})^{2}\,| {\cal C }_{\phi}(A_{1} A_{2})^{2}| \, \right].  \label{Lu2a}
\end{align}
This is another generalization of the uncertainty relation RS. It seems to be easier to use in calculations and practical applications. The related generalization of HR uncertainty relation looks as follows,
\begin{align}
(\Delta_{\phi}A_{1})^{2} &\cdot (\Delta_{\phi}A_{2})^{2} \cdot (\Delta_{\phi}A_{3})^{2} \geq \nonumber \\
&  \frac{1}{6} \left[\, (\Delta_{\phi}A_{1} )^{2}\, \left|\langle [A_{2},A_{3}]\rangle_{\phi}\right|^{2} \right. \nonumber \\
& + \;\;(\Delta_{\phi}A_{2})^{2}\, \left|\langle [A_{1},A_{3}]\rangle_{\phi}\right|^{2}   \nonumber \\
&  + \;\left.  (\Delta_{\phi}A_{3})^{2}\, \left|\langle [A_{1},A_{2}]\rangle_{\phi}\right|^{2} \, \right].  \label{Lu2b}
\end{align}
This last inequality follows from (\ref{Lu2a}) after using all the assumptions and repeating all the steps leading to (\ref{HR2}), (\ref{HR3}),  (\ref{HR4}) and (\ref{B2b}).

Let us now examine simple examples showing how generalized uncertainty relations (\ref{B1a}), (\ref{Lu1a}) work in certain cases. For example, let the state $|\phi\rangle$  be a such in the weak inequality (\ref{RS2}) we have the equality
$ (\Delta_{\phi} A_{1} )\, \cdot\,( \Delta_{\phi} A_{2}) = |{\cal C}_{\phi}(A_{1},A_{2})| $.

The equality in (\ref{RS2})
happens when $|\phi\rangle$ is the solution of the equation
 \begin{equation}
 \delta_{\phi}A_{i}|\phi\rangle = z \delta_{\phi}A_{j}|\phi\rangle, \label{i1}
 \end{equation}
 where $z$ is a complex number
 and  the state vectors $|\phi\rangle$ that are solutions to this equation are called intelligent states (see, e.g., \cite{Jac,fic1,Sh}).

Substituting $|{\cal C}_{\phi}(A_{1},A_{2})| =  (\Delta_{\phi} A_{1} )\, \cdot\,( \Delta_{\phi} A_{2})$
into the left--hand side of inequality (\ref{B1a}) yields exactly inequality (\ref{B2a}), i. e. the weaker version of (\ref{B1a}). A more interesting result is obtained by replacing the product $ (\Delta_{\phi} A_{1} )\, \cdot\,( \Delta_{\phi} A_{2})$ by $|{\cal C}_{\phi}(A_{1},A_{2})| $ on the left-hand side of (\ref{B1a}). We then obtain that
\begin{equation}
   (\Delta_{\phi}A_{3})^{2}\;| {\cal C}_{\phi}(A_{1} A_{2}) \geq  \left|{\cal C}_{\phi}(A_{1} A_{3})\,{\cal C}_{\phi}(A_{3} A_{2})\right|, \label{B3a}
\end{equation}
which gives a lower bound only on the variance $ (\Delta_{\phi}A_{3})^{2}$.

In turn, assume  again that   $|{\cal C}_{\phi}(A_{1},A_{2})| = (\Delta_{\phi} A_{1} ) \cdot ( \Delta_{\phi} A_{2})$ and replace  $|{\cal C}_{\phi}(A_{1},A_{2})|$ in (\ref{Lu1a}) by the product $(\Delta_{\phi} A_{1})
\cdot(\Delta_{\phi} A_{2})$.
This replacement causes the inequality (\ref{Lu1a}) to be transformed into the following:
\begin{equation}
\left(\Delta_{\phi}A_{1}\cdot|{\cal C}_{\phi}(A_{2},A_{3})|-  \Delta_{\phi}A_{2}\cdot|{\cal C}_{\phi}(A_{1},A_{3})|\right)^{2}\leq0. \label{Lu3a}
\end{equation}
The inequality (\ref{Lu3a}) can only
be satisfied if
\begin{equation}
\Delta_{\phi}A_{1}\,\cdot\,|{\cal C}_{\phi}(A_{2},A_{3})|\,\equiv \,\Delta_{\phi}A_{2}\cdot\,|{\cal C}_{\phi}(A_{1},A_{3})|. \label{Lu3b}
\end{equation}
The condition (\ref{Lu3b}) tells us that in a system where we study three non--commuting observables, $A_{1}, A_{2}, A_{3}$,  in a state $|\phi\rangle$ that minimizes the uncertainty principle (\ref{RS2}), for two of them, say $A_{1}, A_{2}$, not all values of ${\cal C}_{\phi}(A_{1},A_{3})$ and ${\cal C}_{\phi}(A_{2},A_{3}) $ are allowed.

\section{Generalizations of sum uncertainty relations}

We will now take Jensen's inequalities (\ref{J2a}) and (\ref{J3a}) and apply them to generalize the "sum uncertainty relations" for two non-commuting observables to the case of $N \geq 3 $ non--commuting observables.
We will use the same assumptions and definitions that were used in the previous Section to derive the generalizations of the uncertainty relations HR and RS.
First, note that from definitions (\ref{dAi}) and (\ref{dF})  it follows that
\begin{equation}
\sum_{i=1}^{N} |\psi_{i}\rangle =  \sum_{i=1}^{N} \delta_{\phi}A_{i}|\phi\rangle \equiv  \delta_{\phi} \left(  \sum_{i=1}^{N} A_{i}\right)|\phi\rangle, \label{sum1}
\end{equation}
and hence  that
\begin{equation}
\left\|\sum_{i=1}^{N} |\psi_{i}\rangle \right\| \equiv \left\| \delta_{\phi} \left(  \sum_{i=1}^{N} A_{i}\right)|\phi\rangle  \right\|  \equiv \Delta_{\phi}( \sum_{i=1}^{N} A_{i}). \label{DF-N}
\end{equation}
Here, the definition (\ref{DF}) is used.

To begin with, let's assume that $N=2$ and use the inequality (\ref{J2a}).
Substituting $|\psi_{1}\rangle$ and $|\psi_{2}\rangle$ into it in the form (\ref{dAi}) and then using (\ref{DF}), (\ref{dF}) and (\ref{DF-N}), we get:
$\Delta_{\phi}A_{1} + \Delta_{\phi}A_{2} \geq \Delta_{\phi}(A_{1} + A_{2})$,
which is the so--called "sum uncertainty relation" derived in \cite{Pat} and mentioned in Sec. 1 (see (\ref{Pat1}).
Repeating all the steps leading to the above relation inequality (\ref{J3a}) gives us the so--called "stronger sum uncertainty relation" (\ref{M12}) discussed in \cite{Mac}.

Now let us assume that we have $ N$ non--commuting observables and use the inequality (\ref{J2a}). Proceeding analogously as in the case of $ N=2$ and using the same assumptions and definitions that allowed us to derive the relation (\ref{Pat1}), we obtain that
\begin{equation}
\sum_{i=1}^{N} \Delta_{\phi}A_{i} \geq \Delta_{\phi}(\sum_{i=1}^{N} A_{i}). \label{Pat2}
\end{equation}
This generalization of the sum uncertainty relation was also discussed in \cite{Pat}. The identical procedure with the same assumptions applied to the Jensen's inequality (\ref{J3a}) gives us
\begin{equation}
\sum_{i=1}^{N} (\Delta_{\phi}A_{i})^{2}\, \geq \, \frac{1}{N}\,\left[\Delta_{\phi}(\sum_{i=1}^{N} A_{i})\right]^{2}. \label{J3b+1}
\end{equation}
This is the simplest generalization of the "stronger sum uncertainty relation" (\ref{M12}) to the case of $N$ non--commuting observables (compare, e.g.,  \cite{Bin,Son}).

\section{Critical points of generalized uncertainty relations}

Let us now examine some critical properties of the uncertainty relations derived in the previous Section.
First, assume that vector $|\phi\rangle$ is an eigenvector of observable $A_{j}$: $|\phi \rangle = |\phi_{a_{j}}\rangle$ and $A_{j} |\phi_{a_{j}}\rangle = a_{j}|\phi_{a_{j}}\rangle$, ($j \leq N$).
The consequence of this assumption is that $\delta_{\phi_{a_{j}}} A_{j} |\phi_{a_{j}}\rangle = 0$, (see Eq. (\ref{dF})).
From Eq.  (\ref{DF}) we conclude that in this case $\Delta_{\phi_{a_{j}}}A_{j} =0$, and from definition (\ref{C}) it follows that ${\cal C}_{\phi_{a_{j}}}(A_{k}A_{j})=0$ for every $ k \neq j \leq N$.
This means that the uncertainty relations, both HR and RS (\ref{RS2}), (\ref{HR2}), and their generalizations (\ref{RS3}) ---  (\ref{Lu2b}) become trivial, i.e. in the case under consideration both their left and right sides become equal to zero.

Let us consider now the sum uncertainty relations and their generalizations.
There is a difference between the cases of two non-commuting observables and $N \geq 3$ non--commuting observables.
To see this we need to  use the property that
\begin{equation}
\delta_{\phi} \left(  \sum_{i=1}^{N} A_{i} \right)|\phi\rangle = \delta_{\phi}  \left(  \sum_{i=1,i\neq j}^{N} A_{i}|\phi\rangle\right) + \delta_{\phi}  A_{j}|\phi\rangle
\label{sum1a}
\end{equation}
which follows from (\ref{sum1}).
Further, if $|\phi \rangle = |\phi_{a_{j}}\rangle$ and $ |\phi_{a_{j}}\rangle$ is an eigenvector of $A_{j}$, ($j \leq N$), then simply
\begin{eqnarray}
\Delta_{\phi_{a_{j}}}( \sum_{i=1}^{N} A_{i}) &\equiv & \left\| \delta_{\phi_{a_{j}}} \left(  \sum_{i=1}^{N} A_{i}\right)|\phi_{a_{j}}\rangle  \right\| \nonumber \\
&\equiv & \left\| \delta_{\phi_{a_{j}}} \left(  \sum_{i=1,i\neq j}^{N} A_{i}\right)|\phi_{a_{j}}\rangle  \right\| \nonumber \\
& \equiv & \Delta_{\phi_{a_{j}}}( \sum_{i=1,i\neq j}^{N} A_{i}). \label{DF-N1a}
\end{eqnarray}
Now suppose that $N=2$ and $j=2$, that is that $A_{2}|\phi_{a_{2}}\rangle = a_{2}|\phi_{a_{2}}\rangle$, then from (\ref{DF-N1a}) it follows that
\begin{equation}
\Delta_{\phi_{a_{2}}}(A_{1} + A_{2}) \equiv \Delta_{\phi_{a_{2}}} A_{1} \label{fi=a2}
\end{equation}
and $\Delta_{\phi_{a_{2}}} A_{2} =0$.
This property means that in the case under consideration, the sum uncertainty relations (\ref{Pat1}) and (\ref{M12}) become trivial relations in the sense that on their left and right sides there will be the same quantity: $\Delta_{\phi_{a_{2}}} A_{1} $ and nothing more.
In this situation,
the inequality (\ref{Pat1}) will take the form  $\Delta_{\phi_{a_{2}}} A_{1} \geq \Delta_{\phi_{a_{2}}} A_{1}$,  whereas the inequality  (\ref{M12}) will take the following form:
$(\Delta_{\phi_{a_{2}}} A_{1})^{2} \geq \frac{1}{2}(\Delta_{\phi_{a_{2}}} A_{1})^{2}$. Both of these inequalities do not provide any  information about lower (or upper) bounds  on $\Delta_{\phi_{a_{2}}} A_{1}$, (see  also \cite{ku1,ku2}).

Let $ N \geq 3$ and $|\phi\rangle = |\phi_{a_{j}}\rangle$, where $j \leq  N$. Then $\Delta_{\phi_{a_{j}}} A_{j}$ = 0, and, as
follows from (\ref{DF-N1a}),  $\Delta_{\phi_{a_{j}}}( \sum_{i=1}^{N} A_{i}) = \Delta_{\phi_{a_{j}}}( \sum_{i=1,i\neq j}^{N} A_{i})$ and thus instead of
uncertainty relation (\ref{Pat2}) we get
\begin{equation}
\sum_{i=1,\, i\neq j}^{N} \Delta_{\phi}A_{i} \geq \Delta_{\phi}(\sum_{i=1,\,i\neq j}^{N} A_{i}). \label{Pat2a}
\end{equation}
Similarly, instead of (\ref{J3b}) we get,
\begin{equation}
\sum_{i=1,\,i \neq j}^{N} (\Delta_{\phi}A_{i})^{2}\, \geq \, \frac{1}{N}\,\left[\Delta_{\phi}(\sum_{i=1, \,i \neq j}^{N} A_{i})\right]^{2}. \label{J3bb}
\end{equation}

Results (\ref{Pat2a}) and (\ref{J3bb}) show that there is certain advantage of the generalized
sum uncertainty relations (\ref{Pat2}) and (\ref{J3b}) over generalized HR and RS uncertainty relations in the case when the vector $|\phi\rangle$ is one of eigenvectors of one of
non--commuting observables: in such a case, sum uncertainty relations allow finding a lower bound on the sum of the remaining non--zero standard deviations or variances, while in such a situation, generalized uncertainty relations  HR and RS do not provide any useful information about the corresponding
bounds.

Let us now analyze another interesting case when the vector $|\phi\rangle$ in (\ref{dAi}) is not an eigenvector of any operators $A_{i}$, $(i = 1,2, \ldots , N)$, but it is such that $|\psi_{j}\rangle = \delta_{\phi}A_{j} |\phi\rangle\;\perp\;
|\psi_{k}\rangle = \delta_{\phi}A_{k} |\phi\rangle,\,(j\neq k)$. Then, as it follows from (\ref{DF}), $\Delta_{\phi}A_{j} > 0$ and $\Delta_{\phi}A_{k} > 0$, but   ${\cal C}_{\phi}(A_{j},A_{k}) \equiv 0$ which  is a direct implication of the definition (\ref{C}).
If $N=2$ and there exists  a vector $|\phi\rangle$ having this property then the right--hand  sides of uncertainty relations (\ref{RS2}) and (\ref{HR2}) are equal to zero but their left--hand sides are non--zero. Therefore the lower bound on the product of standard deviations is equal to zero.
This effect does not occur in this form in the case of generalized uncertainty relations  (\ref{B1a}) --- (\ref{Lu2b}), where $N=3$.
If, for example, $|\psi_{1}\rangle = \delta_{\phi}A_{1} |\phi\rangle\;\perp\;
|\psi_{2}\rangle = \delta_{\phi}A_{2} |\phi\rangle$, then ${\cal C}_{\phi}(A_{1},A_{2}) \equiv 0$ and the right--hand side of (\ref{B1a}) remains unchanged and non--zero,  while the left--hand side of this inequality takes the same form as the left--hand side of the inequality (\ref{B2a}).
In turn, if  $|\psi_{2}\rangle = \delta_{\phi}A_{2} |\phi\rangle\;\perp\;
|\psi_{3}\rangle = \delta_{\phi}A_{3} |\phi\rangle$ (or  $|\psi_{1}\rangle = \delta_{\phi}A_{1} |\phi\rangle\;\perp\;
|\psi_{3}\rangle = \delta_{\phi}A_{3} |\phi\rangle$),  then ${\cal C}_{\phi}(A_{2},A_{3}) \equiv 0$ (or ${\cal C}_{\phi}(A_{1},A_{3}) \equiv 0$),
which causes that
the right hand side of (\ref{B1a}) becomes zero and the left--hand side of this inequality remains unchanged and positive.
In this context,  generalizations of RS uncertainty relation (\ref{Lu1a}), (\ref{Lu2a}) based on the
inequality (\ref{Lu1}), are much more interesting.
Here,  no matter which of the pair of vectors is an orthogonal  pair,  whether $|\psi_{1}\rangle = \delta_{\phi}A_{1} |\phi\rangle\;\perp\;
|\psi_{2}\rangle = \delta_{\phi}A_{2} |\phi\rangle$, or  $|\psi_{1}\rangle = \delta_{\phi}A_{1} |\phi\rangle\;\perp\;
|\psi_{3}\rangle = \delta_{\phi}A_{3} |\phi\rangle$,
or  $|\psi_{2}\rangle = \delta_{\phi}A_{2} |\phi\rangle\;\perp\;
|\psi_{3}\rangle = \delta_{\phi}A_{3} |\phi\rangle$,
on the right side of the inequality (\ref{Lu1a}) we will have
 $\left|{\cal C}_{\phi}(A_{1} A_{2})\,{\cal C}_{\phi}(A_{2} A_{3})\,{\cal C}_{\phi}(A_{1} A_{2})\right| = 0$.
In addition, one of the remaining terms on the right side of this inequality will also be equal to zero, and the other two will be greater than zero, and thus the entire right side of this inequality will be greater than zero.
On the left side of this inequality, there will be no change.
Finally,
on the right side of this inequality, unlike inequalities (\ref{RS3}), (\ref{RS4}), we will obtain non--zero, positive lower bound on the product of three variations $(\Delta_{\phi}A_{1})^{2}\cdot (\Delta_{\phi}A_{2})^{2} \cdot (\Delta_{\phi}A_{3})^{2}$.

Let us now examine the behavior of sum uncertainty relations (\ref{Pat1}), (\ref{M12}) and their generalizations in a similar situation.
If $N=2$ and  $|\psi_{1}\rangle = \delta_{\phi}A_{1} |\phi\rangle\;\perp\;
|\psi_{2}\rangle = \delta_{\phi}A_{2} |\phi\rangle$, then $(\Delta_{\phi}(A_{1} + A_{2}))^{2} \equiv \left\|\delta_{\phi}(A_{1}+A_{2})|\phi\rangle \right\|^{2} =
\left\|\delta_{\phi}A_{1}|\phi\rangle + \delta_{\phi}A_{2}|\phi\rangle \right\|^{2} \equiv \left\|\delta_{\phi}A_{1}|\phi\rangle \right\|^{2} + \left\|\delta_{\phi}A_{2}|\phi\rangle \right\|^{2} \equiv (\Delta_{\phi}A_{1})^{2} + (\Delta_{\phi}A_{2})^{2}$. \linebreak
Using this result and squaring two sides of (\ref{Pat1}) we get that $(\Delta_{\phi}A_{1})^{2} + (\Delta_{\phi}A_{2})^{2}\, + 2\,(\Delta_{\phi}A_{1})\cdot(\Delta_{\phi}A_{2})\geq (\Delta_{\phi}A_{1})^{2} + (\Delta_{\phi}A_{2})^{2}$.
Thus, in the case under consideration, the final form of the inequality (\ref{Pat1}) is $(\Delta_{\phi}A_{1})\cdot(\Delta_{\phi}A_{2})\geq 0$.
This result is the same as for the uncertainty relations RS and HR, obtained under the same assumptions.
In turn, the same reasoning applied to the inequality (\ref{M12}) leads to the result $(\Delta_{\phi}A_{1})^{2} + (\Delta_{\phi}A_{2})^{2} \geq \frac{1}{2}\left( (\Delta_{\phi}A_{1})^{2} + (\Delta_{\phi}A_{2})^{2}\right)$,
which means that in the case under consideration the lower bound on the sum of variances is zero.
Further analysis shows that for $N \geq 3$, the assumption that there exists a vector $|\phi\rangle$ and a pair of non-commuting observables $A_{j}< A_{k},\,(j,k \leq N)$, such that
$|\psi_{j}\rangle = \delta_{\phi}A_{j} |\phi\rangle\;\perp\;
|\psi_{k}\rangle = \delta_{\phi}A_{k} |\phi\rangle,\,(j\neq k)$ does not mean that the right--hand sides of generalized sum uncertainty relations (\ref{Pat2}) and (\ref{J3b}) can be transformed to a trivial form
similar to the case $N = 2$.

The discussion carried out in this Section can be summarized by the observation that
the generalized uncertainty relations for the sum of standard deviations and for the sum of variances are less sensitive to the effects analyzed there than the generalized uncertainty relations HR and RS.


\section{Uncertainty relations and correlations}

The  correlation function
in the large literature is defined as the  matrix element $\langle \phi|\left[\left( A_{i} - \langle A_{i}\rangle_{\phi} \right)  \left( A_{j} - \langle A_{j}\rangle_{\phi}\right)\right]|\phi\rangle$ (see, eg. \cite{Bei,Rei,Roh}).
Simply, according to the definition (\ref{C}) we have:
 $\langle \phi|\left[\left( A_{i} - \langle A_{i}\rangle_{\phi} \right)  \left( A_{j} - \langle A_{j}\rangle_{\phi}\right)\right]|\phi\rangle \equiv \langle \phi| \delta_{\phi} A_{i}\,\delta_{\phi} A_{j}|\phi\rangle = {\cal C}_{\phi}(A_{i},A_{j})$.  This function is also called covariance.
It should be noted that
in some papers the covariance is defined as the real part of ${\cal C}_{\phi}(A_{i},A_{j})$. That is as:  ${\rm cov}_{\phi}(A_{i},A_{j}) = \Re\,[{\cal C}_{\phi}(A_{i},A_{j})]$ (see, e. g.,  \cite{Mac1,Deb}).
However, we should be aware that
 the function ${\rm cov}_{\phi}(A_{i},A_{j})$ is just the classical part of the quantum version of the covariance ${\cal C}_{\phi}(A_{i},A_{j}) $ and does not describe all properties of quantum systems (see, e.g. \cite{ku1}).
The useful formal properties of ${\cal C}_{\phi}(A_{i},A_{j}) $ are:
\begin{eqnarray}
{\cal C}_{\phi}(A_{i},A_{i})& =& (\Delta_{\phi}A_{i})^{2}, \nonumber \\
{\cal C}_{\phi}(A_{i},A_{j})& =& [{\cal C}_{\phi}(A_{j},A_{i})]^{\ast}, \nonumber \\
{\cal C}_{\phi}(A_{i},A_{j} +A_{k}) &=& {\cal C}_{\phi}(A_{i},A_{j})+ {\cal C}_{\phi}(A_{i},A_{k}),\label{cov3}
\end{eqnarray}
There is
${\cal C}_{\phi}(A_{i},A_{j}) = 0 $ if $|\phi\rangle$  is an eigenvector of the operator $A_{i}$ or $A_{j}$.
If the state vector $|\phi \rangle \in {\cal H}$  is not an eigenvector of either observables $A_{i}$ nor $A_{j}$
and ${\cal C}_{\phi}(A_{i},A_{j}) = 0 $, then observables $A_{i}$ and $A_{j}$ are fully uncorrelated in that state.

It is worth noting here that in mathematical statistics an inequality analogous to (\ref{RS2}) is considered, and it is interpreted as an upper bound on the covariance (i.e. the correlation function): People using mathematical statistics know that the covariance is bounded by the product of the standard deviations (see, e.g. \cite{gri} and  \cite{pri}).
Therefore, we can say that inequality (\ref{RS2}) has two faces: Physicists see it as the RS uncertainty relation (i.e., as the lower bound on the product of standard deviations or variances), while statistical mathematicians see it as the upper bound on the covariance (the correlation function).

\begin{theorem}
\label{th1}
For any pair of non--commuting observables $A_{j}$ and  $A_{k}$, ($j \neq k$), acting in a  two--dimensional state space ${\cal H}$, there is no such a state $|\phi\rangle \in {\cal H}$ that is not an eigenvector of either of the observables $A_{j}$ and $A_{k}$, in which
these observables are fully uncorrelated.\
\end{theorem}

\noindent
{\bf Proof:}
Let us assume  that in a two--dimensional state space there exists a state vector $|\phi\rangle \neq 0$  such that the non--commuting observables $A_{j}$ and $A_{k}$, ($j \neq k$) acting in this space are fully uncorrelated in this state,  i.e.,  that  ${\cal C}_{\phi}(A_{j},A_{k}) = 0 $. Assume also that $|\phi\rangle$  is not an eigenvector of either of the observables $A_{j}$ and $A_{k}$. By definitions (\ref{C}) and (\ref{dAi}) these assumptions  mean that $\delta_{\phi}A_{j}|\phi\rangle \equiv |\psi_{j}\rangle \,\perp\, |\psi_{k}\rangle \equiv \delta_{\phi}A_{k}|\phi\rangle$ and by (\ref{dF}) that $|\psi_{j}\rangle \neq 0,\;|\psi_{k}\rangle \neq 0$.
Thus, vectors $|\psi_{j}\rangle$ and $|\psi_{k}\rangle$ can be used as a basis in this space and any other vector from this space, including vector $|\phi\rangle$, can be written in this basis as their linear combination.
We can therefore assume that $|\phi\rangle = \alpha|\psi_{j}\rangle + \beta |\psi_{k}\rangle$.
Hence,  we find that $ \langle\psi_{j}|\phi\rangle = \alpha\,\|\,|\psi_{j}\rangle \|^{2} \neq 0$ and $ \langle\psi_{k}|\phi\rangle = \beta\,\|\,|\psi_{j}\rangle \|^{2} \neq 0$.
From definitions (2), (11), and our assumption that  ${\cal C}_{\phi}(A_{j},A_{k}) = 0 $, it follows that $\langle \phi|\psi_{j}\rangle = \langle \phi|\psi_{k}\rangle \equiv 0$.
Only the coefficients $\alpha = \beta = 0$ are consistent with these results.
This is true because definitions (\ref{dF}), (\ref{dAi}), and our assumption imply that $\langle \phi|\delta_{\phi}A_{j}|\phi\rangle = \langle \phi|\delta_{\phi}A_{k}|\phi\rangle \equiv 0$.
This implies that, contrary to our assumption, the only state vector in a two--dimensional state space in which the non--commuting observables $A_{j}$ and $A_{k}$, ($j \neq k$), could be completely uncorrelated is the vector $|\phi \rangle =0$.$\;\;\Box$\\

\noindent
{\bf Corollary 1:}
In a two--dimensional state space, every pair $A_{1}$ and $A_{2}$ of non--commuting observables acting in that space is correlated in each of the states in that space that are not eigenvectors of operators $A_{1}$ or $A_{2}$.
In other words, in a two--dimensional state space, any quantum state $|\phi\rangle$ that is not an eigenstate of any of the non--commuting observables $A_{1}$ or $A_{2}$ is $(A_{1}A_{2})$--entanglement.\\

\hfill\\
{\bf Definition 1:} We will say the a quantum state $|\phi\rangle$  is
(AB)--entanglement iff correlation of the observables $A$ and $B$ with respect to this state is not factorizable
\begin{equation}
\langle AB\rangle_{\phi} \neq \langle A\rangle_{\phi}\;\langle B\rangle_{\phi}, \label{AB}
\end{equation}
that is iff
\begin{equation}
{\cal C}_{\phi}(A,B) \neq 0. \label{AB1}
\end{equation}

We adopted here the definition used by Krennikov and Basieva, (see \cite{Khr}, Sec. 1).\\

If the system is in a state $|\phi\rangle$ such that $\Delta_{\phi}A_{i} > 0$ and $\Delta_{\phi}A_{j} >0$, then
we can define the following quantity, useful in some applications,
\begin{equation}
r_{\phi}(A_{i},A_{j}) \stackrel{\rm def}{=}  \frac{\left|{\cal C}_{\phi}(A_{i},A_{j}) \right|}{\Delta_{\phi} A_{i} \cdot \Delta_{\phi} A_{j}}, \label{r1}
\end{equation}
which can be considered as  a quantum variant  of  Pearson's  coefficient, i.e. the quantum modification of the correlation coefficient (see, e.g. \cite{Rei,Mac1,Jeb} and also \cite{gri,pri}). There is,
\begin{align}
& r_{\phi}(A_{i},A_{j}) \geq 0,  \;\; r_{\phi}(A_{i},A_{i}) =1, \nonumber \\
&r_{\phi}(A_{i},A_{j}) = r_{\phi}(A_{j},A_{i}). \label{r-2}
\end{align}
If observables $A_{i}$ and $A_{j}$ are fully uncorrelated in the state $|\phi\rangle$ then $r_{\phi}(A_{i},A_{j}) = 0$.

Using the
coefficient  $r_{\phi}(A_{i},A_{j}) $ defined by the formula (\ref{r1})
we can write the uncertainty relation RS (\ref{RS2}) in an equivalent form as
\begin{equation}
r_{\phi}(A_{i},A_{j})  \leq 1. \label{r1-RS}
\end{equation}
This relation  is valid when $\Delta_{\phi}A_{i} > 0$ and $\Delta_{\phi}A_{j} >0$
and we will use this form of (\ref{RS2}) in the remainder of this Section.

It should be noted that $r_{\phi}(A_{i},A_{i}) = 1 $ even though $[A_{i},A_{i}]$=0, which means that the right side of the HR relation (\ref{R1}) is equal to zero.
Moreover,  it may happen that
$r_{\phi}(A_{i},A_{j}) = 0 $ even though $\Delta_{\phi} A_{i} >  0$,  $\Delta_{\phi} A_{j} > 0$.
This happens if there exists such a state $|\phi\rangle$  that
 $\delta_{\phi} A_{i}|\phi\rangle \perp \delta_{\phi} A_{j}|\phi\rangle$ despite the fact that $[A_{i},A_{j}] \neq 0$.
The result $r_{\phi}(A_{i},A_{j}) = 0 $ means that  observables $ A_{i}$ and $A_{j}$ are uncorrelated in the state $|\phi\rangle$. However, this does not exclude that in a state $|\psi\rangle \neq |\phi\rangle$ there may be $r_{\psi}(A_{i},A_{j}) > 0$ and then the value $r_{\psi}(A_{i},A_{j})$ describes the size of a correlation.
The case $r_{\phi}(A_{i},A_{j}) =1 $ occurs when  the equality holds in the relations (\ref{RS2})  and describes fully correlated observables $ A_{i}$  and $A_{j}$ in the state   $|\phi\rangle$.
The equality in (\ref{RS2}), or the case $r_{\phi}(A_{i},A_{j}) =1 $, happens when $|\phi\rangle$ is the solution of the equation
(\ref{i1}).
So, we can say that
the non--commuting observables $A_{i}$ and $A_{j}$ are fully correlated  in the state  $|\phi\rangle$ solving the Eq. (\ref{i1}), i.e., in the intelligent states of the system, or that the state $|\phi\rangle$ is an intelligent state for non--commuting observables $A_{i}$ and $A_{j}$ when  $r_{\phi}(A_{i},A_{j}) =1 $.

Now suppose that $|\phi\rangle$  is not an eigenvector of any of the operators $\{ A_{j} \}_{j = 1}^{N}$ and that $[A_{j},A_{k}] \neq 0$ if $j\neq k$. Then (see (\ref{dAi}) and (\ref{DF})), $\delta_{\phi}A_{i}|\phi\rangle \neq 0$ and $\Delta_{\phi}A_{i}  > 0$ for all $i = 1,2, \ldots, N$.
Using these assumptions we can divide the inequalities (\ref{RS3}),  (\ref{RS4}), (\ref{B1a}), (\ref{B2a}), (\ref{Lu1a}) and (\ref{Lu2a}) by their left--hand sides and recalling the definition (\ref{r1}) write them in an equivalent way as the inequalities depending only on  $  r_{\phi}(A_{i},A_{j})$.
And so from (\ref{RS3}) we get
\begin{equation}
r_{\phi}(A_{1},A_{2})   \cdot  r_{\phi}(A_{2},A_{3})  \cdot   r_{\phi}(A_{1},A_{3}) \,\leq \,1. \label{RS3r}
\end{equation}
In turn, instead of (\ref{RS4}), we get
\begin{align}
&r_{\phi}(A_{1},A_{2}) \cdot  r_{\phi}(A_{1},A_{3}) \cdot  r_{\phi}(A_{1},A_{4}) \cdot \nonumber \\
& r_{\phi}(A_{2},A_{3}) 
 \cdot  r_{\phi}(A_{2},A_{4}) \cdot
r_{\phi}(A_{3},A_{4})\,\leq \,1. \label{RS4r}
\end{align}
In a similar way, using the equivalent form (\ref{r1-RS}) of the RS uncertainty relation (\ref{RS2}), we can generalize this relation to the case of $N$ non--commuting observables. This will simply be the product of an appropriate number of functions $  r_{\phi}(A_{i},A_{j})$ containing all possible combinations of pairs of observables  $A_{i}$ and  $A_{j}$
with respect to the properties (\ref{r-2}).

Then, the same method applied to (\ref{B1a}) gives us the following inequality,
\begin{equation}
1 + r_{\phi}(A_{1},A_{2})\; \geq \;2\, r_{\phi}(A_{1},A_{3}) \cdot  r_{\phi}(A_{2},A_{3}), \label{B1ar}
\end{equation}
and instead of (\ref{B2a}),  we obtain the obvious result that was to be expected in light of the above observations:
\begin{equation}
 r_{\phi}(A_{1},A_{3}) \cdot  r_{\phi}(A_{2},A_{3})\, \leq \, 1. \label{B2ar}
\end{equation}

Analyzing the uncertainty relation (\ref{B1a})  resulting from the Buzano inequality (\ref{B1}) we studied in Sec. 3 its implications in the case when the state $|\phi\rangle$ is an intelligent state for the observables $A_{1}$ and $A_{2}$, which corresponds to the assumption that $r_{\phi}(A_{1},A_{2})=1$.
We can  do the same for this inequality written using Pearson coefficients, i.e. in the case of (\ref{B1ar}), and substitute  $r_{\phi}(A_{1},A_{2})=1$ into it.
It may be somewhat surprising that such a substitution yields an inequality that coincides with inequality (\ref{B2ar}), i.e. with a weaker version of inequality (\ref{B1ar}).
Let us also note that by substituting  $r_{\phi}(A_{1},A_{2})=1$ into the inequality (\ref{RS3r}) we also obtain exactly the same result, i.e. (\ref{B2ar}).

The same reasoning that allowed us to obtain inequalities (\ref{RS3r}) - (\ref{B2ar}) applied to inequality (\ref{Lu1a})
allows us to rewrite the uncertainty relation  (\ref{Lu1a}) in the following form
\begin{align}
&1 \,+ 2 \,\, r_{\phi}(A_{1},A_{2})\cdot  r_{\phi}(A_{1},A_{3})\cdot  r_{\phi}(A_{2},A_{3})\, \geq \nonumber \\
&\geq  \left(r_{\phi}(A_{1},A_{2})\right)^{2} 
+ \left(r_{\phi}(A_{1},A_{3})\right)^{2} 
+\left(r_{\phi}(A_{2},A_{3})\right)^{2}. \label{Lu1ar}
\end{align}
A similar procedure applied to inequality (\ref{Lu2a}) leads to an inequality that is identical to simply adding the correspondingly written three inequalities (\ref{r1-RS}) side by side.
The uncertainty relation (\ref{Lu1a}) based on the
inequality (\ref{Lu1}) and the relation (\ref{Lu1ar}) are equivalent to each other,
except that the inequality (\ref{Lu1ar}) shows the relationship between the correlation coefficients of the observables $A_{1},  A_{2}$ and $A_{3}$ in state $|\phi\rangle$, or more precisely, the the constraints on the size of their correlation in this state.

The quantum Pearson coefficient (\ref{r1}) and the related inequality (\ref{r1-RS}) equivalent to the RS uncertainty relation  allow the use tools of mathematical statistics  to study correlations in the cases of many non--commuting observables occurring in the quantum system under study.
Correlations of such observables in a given state $|\phi\rangle$ can be studied, for example, using the correlation matrix $\mathbf{C_{M(k)}}, \;(k=2,3,  \dots)$, (see, eg., \cite{Rou,Wang,gna,ho}).
Thus, for two non--commuting observables,  $A_{1}, A_{2}$, the correlation matrix looks as follows (compare, e.g. \cite{Rou}):
\begin{equation}
\mathbf{C_{M(2)}} = \left(
                      \begin{array}{cc}
                        1 & r_{\phi}(A_{1},A_{2}) \\
                        r_{\phi}(A_{1},A_{2}) & 1 \\
                      \end{array}
                    \right). \label{C2}
\end{equation}
Here properties of $r_{\phi}(A_{i},A_{j})$ listed in Eq. (\ref{r-2}) were used.
When we study the correlations of three observables $A_{1}, A_{2}$, and $A_{3}$ in a state $|\phi\rangle$, we must use the correlation matrix in the following form:
\begin{equation}
\mathbf{C_{M(3)}} = \left(
                      \begin{array}{ccc}
                        1 & r_{\phi}(A_{1},A_{2}) & r_{\phi}(A_{1},A_{3}) \\
                        r_{\phi}(A_{1},A_{2}) & 1 & r_{\phi}(A_{2},A_{3}) \\
                        r_{\phi}(A_{1},A_{3}) & r_{\phi}(A_{2},A_{3}) & 1 \\
                      \end{array}
                    \right). \label{C3}
\end{equation}
For more obserwables, the correlation matrix is constructed analogously.

Some information about the correlation of observables in state $|\phi\rangle$ can be obtained by analyzing the determinant ${\bf R}_{k} \stackrel{\rm def}{=} \det{\mathbf{C_{M(k)}}}$ of the correlation matrix.
Values of the determinant of the valid correlation matrix are between 0 and 1. Value 1 occurs when all observables are perfectly uncorrelated. Near zero value of the determinant means that observables are highly correlated with each other.
And so, for the  matrix $\mathbf{C_{M(2)}} $ we get,
\begin{equation}
{\bf R}_{2} =  \det{\mathbf{C_{M(2)}}} = 1 - \left(\,r_{\phi}(A_{1},A_{2})\,\right)^{2}\;\geq\;0, \label{R2}
\end{equation}
which is strictly equivalent to inequality (\ref{r1-RS}), i.e., the modified RS uncertainty relation.

Let us now consider the case of three observables.
The determinant of the correlation matrix $\mathbf{C_{M(3)}} $ is equal to,
\begin{align}
{\bf R}_{3} & =  \det{\mathbf{C_{M(3)}}}  =  1\, + \nonumber \\
&+ 2 \,r_{\phi}(A_{1},A_{2})\;r_{\phi}(A_{2},A_{3})\;r_{\phi}(A_{1},A_{3}) \nonumber \\
&-\left(\,r_{\phi}(A_{1},A_{2})\,\right)^{2}-\left(\,r_{\phi}(A_{1},A_{3})\,\right)^{2} \nonumber \\
&-\;\left(\,r_{\phi}(A_{2},A_{3})\,\right)^{2}\; \geq 0, \label{R3}
\end{align}
and its non--negativity results from the  generalized uncertainty relation (\ref{Lu1ar}), which is a consequence
of the inequality (\ref{Lu1}).
The above--men\-tio\-ned and used tools of  mathematical statistics, when applied to the case of 4 or more observables, allow us to find uncertainty relations of a similar type to that derived from the generalization of the Schwarz inequality (\ref{Lu1}) described by Lupu and Schwarz \cite{Lu}.

At this point, it would be appropriate to recall the Definition 1, (\ref{AB}), (\ref{AB1}) and generalize it as follows:\\
\hfill\\
{\bf Definition 2:} Let us consider three non--commuting observables  $A, B$ and $C$. We will say the a quantum state $|\phi\rangle$  is
(ABC)--entanglement iff it is simultaneously (AB)--entanglement,  (AC)--entanglement and (BC)--entaglement.\\

Let us now return to the analysis of the uncertainty relation (\ref{Lu1a}) resulting from the inequality (\ref{Lu1}) and the consequences of the assumption that $|\phi\rangle$ is an intelligent state for the observables $A_{1}$ and $A_{2}$, which led us to some constraints in the form of equation (\ref{Lu3b}).
Thus, let us assume that $|\phi\rangle$ is the intelligent state for non--commuting observables $A_{1}$ and $A_{2}$, i. e., that   $r_{\phi}(A_{1},A_{2})=1$, and apply this assumption to equation (\ref{R3})
as a result, we obtain the following inequality,
\begin{equation}
{\bf R}_{3} =  \det{\mathbf{C_{M(3)}}}  \equiv -\left[r_{\phi}(A_{1},A_{3})-r_{\phi}(A_{2},A_{3})\right]^{2}\,\geq 0.
 \label{R3r=1}
\end{equation}
The only acceptable, selfconsistent solution to this weak inequality is
\begin{equation}
r_{\phi}(A_{1},A_{3})\;\equiv \;r_{\phi}(A_{2},A_{3}). \label{R3r=1-sol}
\end{equation}
This condition is much easier to interpret than the previously obtained corresponding condition (\ref{Lu3b}).
 Simply, condition (\ref{R3r=1-sol}) states that if non--commuting observables $A_{1}, A_{2}$, and $A_{3}$ are in the state $|\phi\rangle$, which is an intelligent state for observables $A_{1}$  and $A_{2}$, then the size of the correlations of observables $A_{1}$ and $A_{3}$, and $A_{2}$ and $A_{3}$ in this state, must be the same. In other words, we have proved the following theorem:
\begin{theorem}
\label{th2}
Let $A_{1}, A_{2}$, and $A_{3}$ be non--commuting observables acting in the quantum system under study.
If this system is in the state $|\phi\rangle$, which is an intelligent state for the pair of observables $A_{1}$ and $A_{2}$, i.e., if, $r_{\phi}(A_{1},A_{2})=1$, then the correlation size of the pair $A_{1}$ and $A_{3}$ and of the pair $A_{2}$ and $A_{3}$ in this state must be the same: $r_{\phi}(A_{1},A_{3})\;\equiv \;r_{\phi}(A_{2},A_{3})$.
\end{theorem}

So if the quantum system is in the state $|\phi\rangle$, which is an intelligent state for two of the three non--commuting observables, $A_{1}, A_{2}, A_{3}$  say, for $A_{1}$ and $A_{2}$, then it is enough to find  $r_{\phi}(A_{1},A_{3})$ (or $r_{\phi}(A_{2},A_{3})$) to know the size of $r_{\phi}(A_{2},A_{3})$ (or $r_{\phi}(A_{1},A_{3})$.
By Definition 2, such a state $|\phi\rangle$ is ($A_{1}A_{2}A_{3}$)--entanglement.

Let us now examine the possible scenarios in the system with three non--commuting observables, $A_{1}, A_{2}, A_{3}$ resulting from the inequalities (\ref{Lu1a}),  (\ref{Lu1ar}) and (\ref{R3}).
First, let us assume that observables $A_{1}$ and $A_{2}$ are  uncorrelated in state $|\phi\rangle$, that is that ${\cal C}_{\phi}(A_{1},A_{2}) = 0$, and therefore that $r_{\phi}(A_{1},A_{2}) =0$.
Then the relations (\ref{Lu1ar}), (\ref{R3})  take the form
\begin{equation}
1\, -\, \left(\,r_{\phi}(A_{1},A_{3})\,\right)^{2} \,-\,\left(\,r_{\phi}(A_{2},A_{3})\,\right)^{2}\; \geq 0, \label{r-A1A2=0}
\end{equation}
which means that, e. g., the upper bound for $\left(\,r_{\phi}(A_{2},A_{3})\,\right)^{2}$ has the form $1\, -\, \left(\,r_{\phi}(A_{1},A_{3})\,\right)^{2} \,\geq\,\left(\,r_{\phi}(A_{2},A_{3})\,\right)^{2}$.
This situation is presented graphically in Fig.  \ref{r-12-a}(a).
In turn, when $r_{\phi}(A_{1},A_{2}) =0$ and $r_{\phi}(A_{1},A_{3}) =0$, then relations (\ref{Lu1ar}) and  (\ref{R3})   reduce to the relations (\ref{r1-RS}), (\ref{R2}), describing the case of two non-commuting observables:
$1 - \left(\,r_{\phi}(A_{2},A_{3})\,\right)^{2}\;\geq\;0$.

In the general case of a quantum system in which we study three non--commuting observables,  $A_{1}, A_{2}, A_{3}$, the allowed values of the Pearson coefficient $r_{\phi}(A_{1},A_{3})$ and $r_{\phi}(A_{2},A_{3})$
(and thus the size of correlations of observables $A_{1}, A_{3}$ and $A_{2},A_{3}$ in the state $|\phi\rangle$) for a given values of $r_{\phi}(A_{1},A_{2})$ are presented in Figs \ref{r-12-a} and \ref{r-12-b}.
The most interesting picture can be observed when the state $|\phi\rangle$ is close to the ideal state for the observables $A_{1}$ and $A_{2}$ , for which $r_{\phi}(A_{1},A_{2}) = 1$

\pagebreak

\begin{widetext}

\begin{figure}[H]
    \centering
    \begin{subfigure}[b]{0.40\textwidth}
        \centering
        \includegraphics[width=\textwidth]{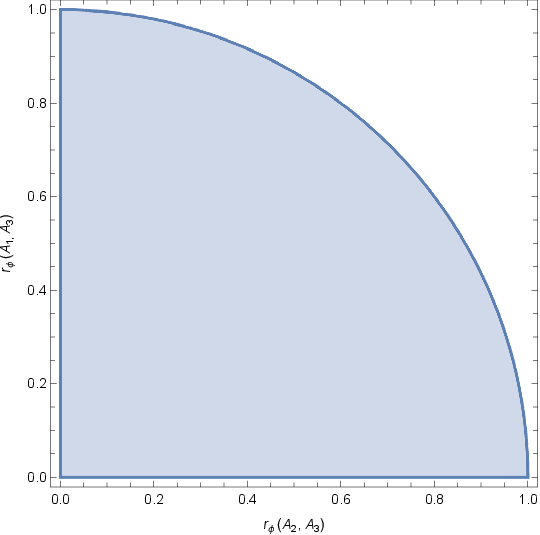}
        \caption{$r_{\phi}(A_{1},A_{2})=0,000$}
        \label{r-a}
    \end{subfigure}
    \hfill
    \begin{subfigure}[b]{0.40\textwidth}
        \centering
        \includegraphics[width=\textwidth]{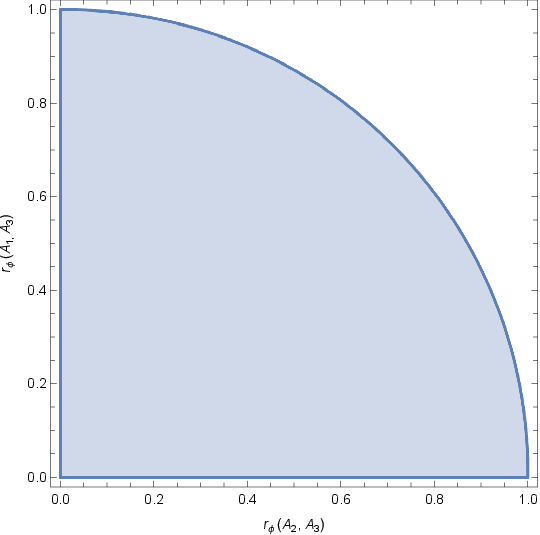}
        \caption{$r_{\phi}(A_{1},A_{2})=0,010$}
        \label{r-b}
    \end{subfigure}
    \\
    \begin{subfigure}[b]{0.40\textwidth}
        \centering
        \includegraphics[width=\textwidth]{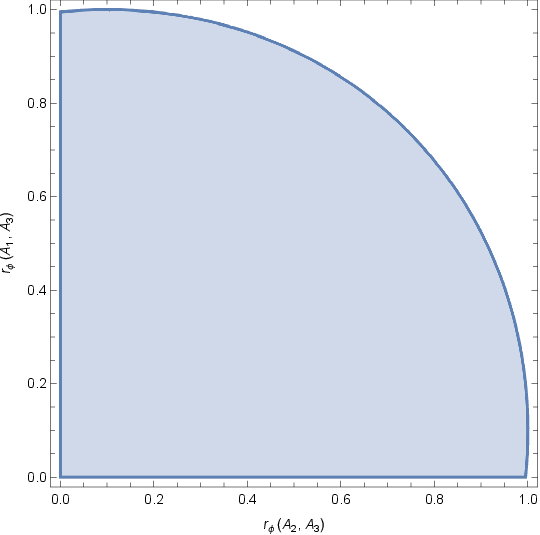}
        \caption{$r_{\phi}(A_{1},A_{2})=0,100$}
        \label{r-c}
    \end{subfigure}
     \hfill
    \begin{subfigure}[b]{0.40\textwidth}
        \centering
        \includegraphics[width=\textwidth]{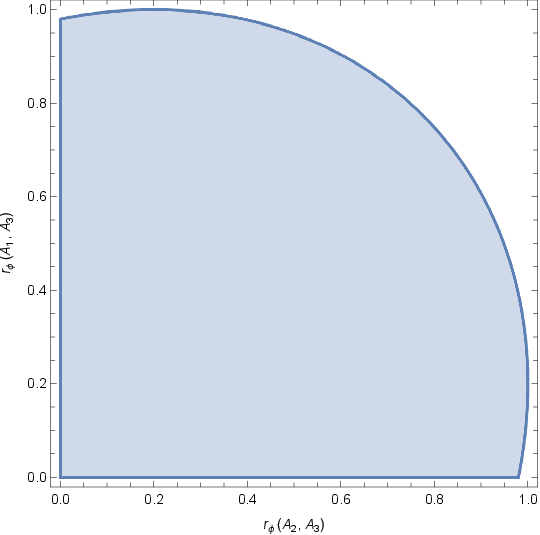}
        \caption{$r_{\phi}(A_{1},A_{2})=0,200$}
        \label{r-d}
    \end{subfigure}\\
     \begin{subfigure}[b]{0.40\textwidth}
        \centering
        \includegraphics[width=\textwidth]{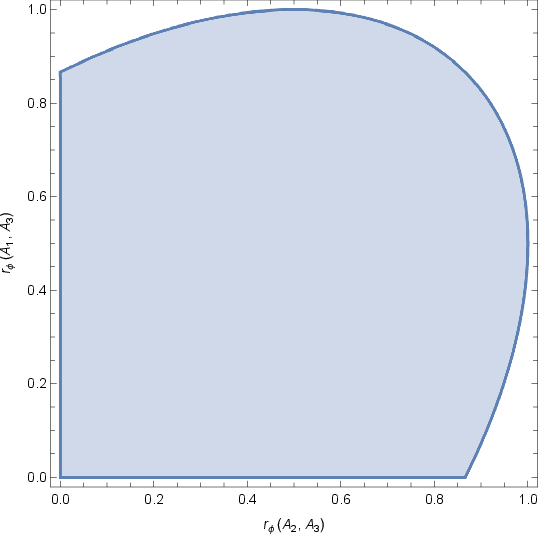}
        \caption{$r_{\phi}(A_{1},A_{2})=0,500$}
        \label{r-e}
    \end{subfigure}
     \hfill
    \begin{subfigure}[b]{0.40\textwidth}
        \centering
        \includegraphics[width=\textwidth]{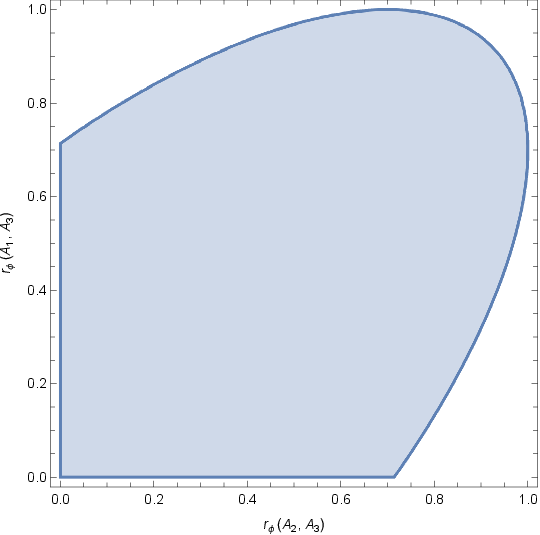}
        \caption{$r_{\phi}(A_{1},A_{2})=0,700$}
        \label{r-f}
    \end{subfigure}
    \caption {Allowable values $r_{\phi}(A_{1},A_{3})$ and $r_{\phi}(A_{2},A_{3})$ in a quantum system for given values $r_{\phi}(A_{1},A_{2})$. }
    \label{r-12-a}
\end{figure}

\pagebreak

\begin{figure}[H]
    \centering
    \begin{subfigure}[b]{0.40\textwidth}
        \centering
        \includegraphics[width=\textwidth]{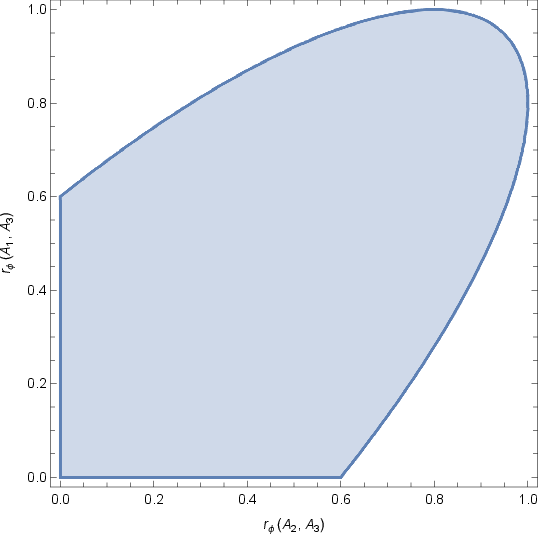}
        \caption{$r_{\phi}(A_{1},A_{2})=0,800$}
        \label{r2-a}
    \end{subfigure}
    \hfill
    \begin{subfigure}[b]{0.40\textwidth}
        \centering
        \includegraphics[width=\textwidth]{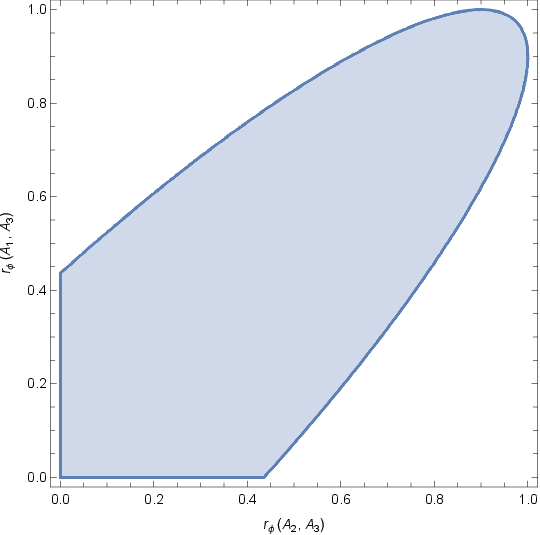}
        \caption{$r_{\phi}(A_{1},A_{2})=0,900$}
        \label{r2-b}
    \end{subfigure}
    \\
    \begin{subfigure}[b]{0.40\textwidth}
        \centering
        \includegraphics[width=\textwidth]{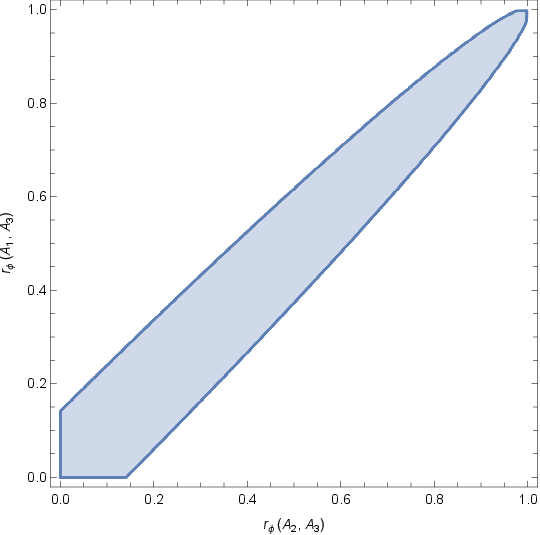}
        \caption{$r_{\phi}(A_{1},A_{2})=0,990$}
        \label{r2-c}
    \end{subfigure}
     \hfill
    \begin{subfigure}[b]{0.40\textwidth}
        \centering
        \includegraphics[width=\textwidth]{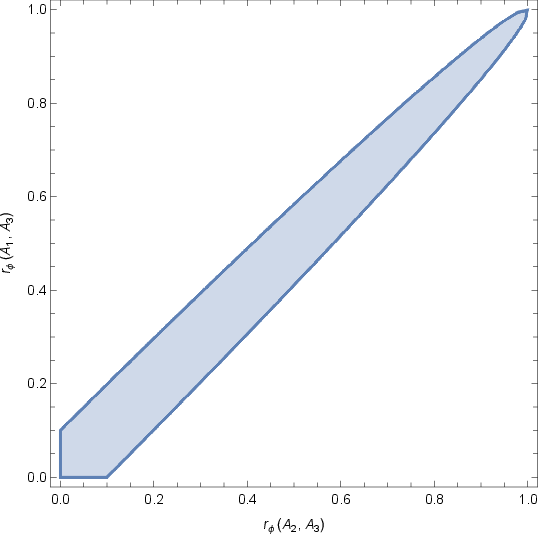}
        \caption{$r_{\phi}(A_{1},A_{2})=0,995$}
        \label{r2-d}
    \end{subfigure}\\
     \begin{subfigure}[b]{0.40\textwidth}
        \centering
        \includegraphics[width=\textwidth]{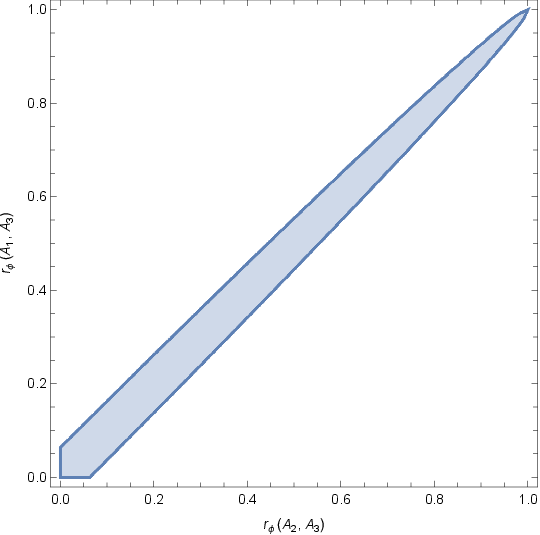}
        \caption{$r_{\phi}(A_{1},A_{2})=0,998$}
        \label{r2-e}
    \end{subfigure}
     \hfill
    \begin{subfigure}[b]{0.40\textwidth}
        \centering
        \includegraphics[width=\textwidth]{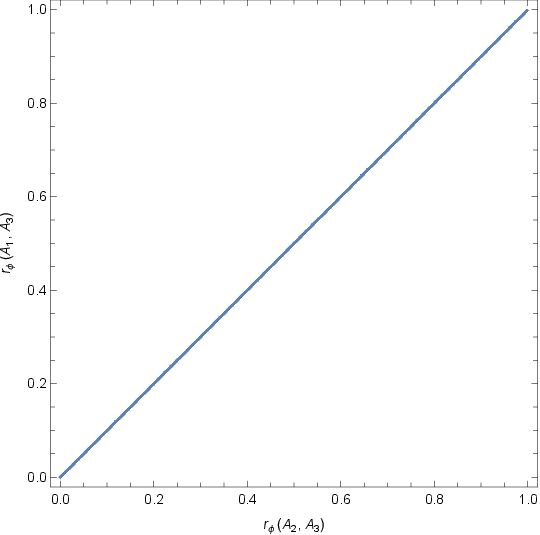}
        \caption{$r_{\phi}(A_{1},A_{2})=1,000$}
        \label{r2-f}
    \end{subfigure}
    \caption {Allowable values $r_{\phi}(A_{1},A_{3})$ and $r_{\phi}(A_{2},A_{3})$ in a quantum system for given values $r_{\phi}(A_{1},A_{2})$. }
    \label{r-12-b}
\end{figure}

\end{widetext}

As can be seen from Figure \ref{r-12-b}, for $r_{\phi}(A_{1},A_{2}) > 0.7$ the region of allowed values of $r_{\phi}(A_{1},A_{3})$ and $r_{\phi}(A_{2},A_{3})$
is increasingly narrower, and the narrowest is for $r_{\phi}(A_{1},A_{2}) = 1$.
This means that for such values of $r_{\phi}(A_{1},A_{2})$ knowing, for example,
$r_{\phi}(A_{1},A_{3})$
one can estimate the allowable values of $r_{\phi}(A_{2},A_{3})$
with high accuracy, and this accuracy will be greater the closer the state $|\phi\rangle$ is to the ideal state for observables $A_{1}$ and $A_{2}$.

The extreme case is shown in Figure \ref{r-12-b}(f): if state $\phi\rangle$ is an ideal state for $A_{1}$ and $A_{2}$, then it is enough to find $r_{\phi}(A_{1},A_{3})$ to know the exact value of $r_{\phi}(A_{2},A_{3})$. In this case, $r_{\phi}(A_{1},A_{3})=r_{\phi}(A_{2},A_{3})$  in full agreement with Theorem 2.

Another conclusion follows from the analysis of relations (\ref{Lu1ar}) and (\ref{R3}): \\
\hfill\\
{\bf Corollary 2:} If $|\phi\rangle$ is simultaneously an intelligent state for observables $A_{1}, A_{2}$ and $A_{2},A_{3}$, (that is if  $ r_{\phi}(A_{1},A_{2}) =r_{\phi}(A_{2},A_{3})=1 $)
then it must also be an intelligent state for observables $A_{1},A_{3}$, (i. e. then it must be $r_{\phi}(A_{1},A_{3})=1$).\\

The discussion in this Section and the inequalities considered here demonstrate the direct connection of the uncertainty relations with the bounds for the correlation function ${\cal C}_{\phi}(A_{i},A_{j})$ and the correlation  coefficient $r_{\phi}(A_{i},A_{j})$. More precisely, it indicates the connection of the uncertainty relations with the size of the correlation of non--commuting observables $A_{i}$ and $A_{j}$ in state $|\phi\rangle$.
It also demonstrates unexpected properties of correlations in a quantum system with three non--commuting observables. These properties are a consequence of the uncertainty relation for these observables, which follows from the generalization of the Schwarz inequality to the case of three vectors.


\section{Final remarks}

In systems exhibiting quantum properties, uncertainty relations determine the possible dispersion of measurement results of physical quantities described by non--commuting operators.
They were discovered by analyzing the measurement of the position and momentum of a particle exhibiting quantum properties \cite{H,H2}, and then they were generalized to two non-commuting observables (see \cite{Robertson,Schrod-1930}). This generalization was derived using the Schwarz inequality. Some time ago, a new formulation of the uncertainty relation appeared, relating two non--commuting observables to the sum of their standard deviations \cite{Pat}, or to the sum of their variances  \cite{Mac}.
Studies have also begun on generalizations of the uncertainty relation to the case of three or more non--commuting observables (see eg. \cite{Bin,Son,Had,sh1} and others).
These kinds of generalizations and extensions of the uncertainty principle to the case of three or more non--commuting observables are based on generalizations of the Schwarz inequality, which is known in functional analysis for two vectors from a space with an inner product, to the case of three or more vectors from such a space, and on extensions of the triangle inequality (including the triangle inequality of the second kind) to many vectors.
All generalizations and extensions of the Schwarz inequality used in this paper, including new ones, have been discussed and presented in Sec 2. This Section also discusses extensions of the triangle inequality based on Jensen's inequality and similar extensions for sums of squared vector norms.
The inequalities analyzed in Sec. 2 are used in Sec. 3 to study generalizations of the Heisenberg--Robertson and Schrodinger--Robertson uncertainty relations for two or more non--commuting observables. In turn, Sec. 4 analyzes generalizations of the so--called sum uncertainty relations to the case of $N \geq 3$ non--commuting observables.
In Sec. 5, critical points of the uncertainty relations analyzed in the previous chapters are examined. In particular, we analyze cases where the system is in a state that is an eigenstate of one of the non--commuting observables. Then the uncertainty relations HR and SR and their generalizations become trivial  because both their sides are zero. In such cases, in turn, the advantage of the generalizations of sum uncertainty relations becomes apparent, because then their left and right sides are nonzero.
Another non--trivial case is when the system is in a state $|\phi\rangle$  that $\delta_{\phi}A_{j} |\phi\rangle\;\perp\;
\delta_{\phi}A_{k} |\phi\rangle,\,(j\neq k)$. Then, for two non-commuting observables, the right-hand sides of the relations HR and SR are equal to zero, while their left-hand sides are non-zero. When the number of non-commuting observables is greater than two, the situation is more complicated and is described in detail in the aforementioned Section. It is also shown therein that
the generalized uncertainty relations for the sum of standard deviations and for the sum of variances are less sensitive to these effects  than the generalized uncertainty relations HR and RS.
The connections of uncertainty relations and generalized uncertainty relations with correlations in a quantum system are analyzed and discussed in detail in Sec. 6.
One of the conclusions resulting from this analysis is the observation that
physicists see the inequality (\ref{RS2}) as the RS uncertainty relation (i.e., as the lower bound on the product of standard deviations or variances), while statistical mathematicians see it as the upper bound on the covariance (the correlation function).
One of the conclusions arising from the considerations in Sec. 6 is the observation that for any pair of non--commuting observables $A_{j}$ and  $A_{k}$ acting in a  two--dimensional state space ${\cal H}$, there is no such a state $|\phi\rangle \in {\cal H}$ that is not an eigenvector of either of the observables $A_{j}$ and $A_{k}$, in which
these observables are fully uncorrelated.
Another conclusion is that
in a two--dimensional state space, any quantum state $|\phi\rangle$ that is not an eigenstate of any of the non--commuting observables $A_{j}$ or $A_{k}$ is $(A_{j}A_{k})$--entangled.
In Sec. 6 the quantum variant of the Pearson coefficient is also analyzed.
Using it, the uncertainty relations analyzed in Sections 3 and 4 are written as inequalities relating Pearson's coefficients for pairs of non--commuting observables (see (\ref{r1}) and the related inequality (\ref{r1-RS})) .
 This allows the use tools of mathematical statistics  to study correlations in the cases of many non--commuting observables occurring in the quantum system under study. 
 Studying  correlations of such observables in a given state $|\phi\rangle$
we can use the correlation matrix $\mathbf{C_{M(k)}}, \;(k=2,3,  \dots)$, (see Sec. 6, formulae (\ref{C2}) and (\ref{C3})).
Having the correlation matrix, we can extract information about the correlations of observables in state $\phi\rangle$ by analyzing the determinant of this matrix (see (\ref{R2}), (\ref{R3})).

In Sec. 6 we also define the notion of ABC--entanglement for a given state $|\phi\rangle$, (where  $ A, B$ and $C$ are non--commuting observables):
We will say the a quantum state $|\phi\rangle$  is
(ABC)--entanglement iff it is simultaneously (AB)--entanglement,  (AC)--entanglement and (BC)--entaglement.
In addition,
we also perform therein a more detailed analysis of the case of 3 non--commuting observables based on the uncertainty relation resulting from the generalization of the Schwarz inequality (\ref{Lu1}) by Lupo and Schwarz \cite{Lu}.
In particular, we show that if this system is in the state $|\phi\rangle$, which is an intelligent state for the pair of observables $A_{1}$ and $A_{2}$, i.e., if, $r_{\phi}(A_{1},A_{2})=1$, then the correlation size of the pair $A_{1}$ and $A_{3}$ and of the pair $A_{2}$ and $A_{3}$ in this state must be the same: $r_{\phi}(A_{1},A_{3})\;\equiv \;r_{\phi}(A_{2},A_{3})$ (see Theorem 2, Sec. 6).
The possible scenarios in the system with three non--commuting observables, $A_{1}, A_{2}, A_{3}$ resulting from the inequalities (\ref{Lu1a}),  (\ref{Lu1ar}) and (\ref{R3}) are also discussed in Sec. 6
An example of this discussion is the analysis of the case when   observables $A_{1}$ and $A_{2}$ are  uncorrelated in state $|\phi\rangle$ (i.e. when  ${\cal C}_{\phi}(A_{1},A_{2}) = 0$, which implies that $r_{\phi}(A_{1},A_{2}) =0$)  and
this situation is presented graphically in Fig.  \ref{r-12-a}(a).
In the general case of three non--commuting observables,  $A_{1}, A_{2}, A_{3}$, the allowed values of the Pearson coefficient $r_{\phi}(A_{1},A_{3})$ and $r_{\phi}(A_{2},A_{3})$
(and thus the size of correlations of observables $A_{1}, A_{3}$ and $A_{2},A_{3}$ in the state $|\phi\rangle$) for a given values of $r_{\phi}(A_{1},A_{2})$ are presented in Figs \ref{r-12-a} and \ref{r-12-b}.
Note that the most interesting picture can be observed when the state $|\phi\rangle$ is close to the ideal state for the observables $A_{1}$ and $A_{2}$ , for which $r_{\phi}(A_{1},A_{2}) = 1$.

To summarize,
we showed how to derive a generalization of the Schwarz inequality for two vectors to the case of three or more vectors and analyzed the properties of other similar generalizations. We also described generalizations of such inequalities, such as various variants of the triangle inequality resulting from Jensen's inequality. Using these generalized inequalities, we derived various variants of the RS uncertainty relation  for two or more non--commuting observables, as well as a generalization of "sum uncertainty relations" to the case of $N \geq 3$ non--commuting observables,  and analyzed the properties of these generalizations.
By examining the connections between the RS uncertainty relation and the correlation function of the given observables in the state of the quantum system under study, we showed that the RS uncertainty relations (including the generalized ones) can be written in an equivalent way using the quantum version of Pearson coefficients.
It seems that all these considerations and results, and in particular those relating to the case of three non--commuting observables,
may be important for quantum metrology, quantum technologies and quantum communication and may find applications there.\\

\begin{acknowledgments}
The author would like to thank Janusz Matkowski for the inspiring discussion.
This work was supported by
the program of the  Ministry
of Science and Higher Education in Poland under the name "Regional
Initiative of Excellence", Project No. RID/SP/0050/2024/1.
\end{acknowledgments}

\section*{Conflict of Interest}
The author declares no conflict of interest.
%
The author declares that there is not any personal, academic interest, or any other factors that may be perceived to influence the objectivity, integrity or value of the study.

\section*{Data Availability}
This manuscript has no
associated data, or the data will not be deposited. [Author's
comment: This is a theoretical work and analytical calculations are made. Therefore, no data are required].

\end{document}